\documentclass[final,5p,times,twocolumn]{elsarticle}

\makeatletter
\def\ps@pprintTitle{%
  \let\@oddhead\@empty
  \let\@evenhead\@empty
  \let\@oddfoot\@empty
  \let\@evenfoot\@empty
}
\makeatother

\usepackage[normalem]{ulem}
\usepackage{cancel}
\usepackage{pdfcomment}

\usepackage{amsmath}
\usepackage{amssymb}
\usepackage{latexsym}
\usepackage{amsthm}
\usepackage{color,soul} 
\usepackage{todonotes}

\usepackage{array}
\usepackage{graphicx}
\graphicspath{{images/}}

\usepackage[font=footnotesize]{caption}
\usepackage[font=footnotesize]{subcaption}

\usepackage{url}

\usepackage{placeins}

\newcommand{\diag}{\operatorname{diag}}
\newcommand{\Tr}{\operatorname{Tr}}
\renewcommand{\Re}{\operatorname{Re}}
\renewcommand{\Im}{\operatorname{Im}}

\theoremstyle{plain}   

\newtheorem{lemma}{Lemma}

\newtheorem{observation}{Observation}

\begin{document}

\begin{frontmatter}

\title{Compensating Coarse Quantization in Massive MIMO: Channel Estimation and BER under Imperfect CSI}

\author[sussex]{Reza~Mohammadkhani\corref{cor1}}
\ead{mohammadkhani@sussex.ac.uk}
\author[iau]{Azad~Azizzadeh}
\author[razi]{Seyed~Vahab~Al-Din~Makki}
\author[edinburgh]{John~Thompson}
\author[sussex]{Maziar~Nekovee}

\cortext[cor1]{Corresponding author: R. Mohammadkhani.}

\address[sussex]{School of Engineering and Informatics, University of Sussex, UK}
\address[iau]{Electrical and Computer Engineering Department, IAU, Saqqez Branch, Iran}
\address[razi]{Electrical Engineering Department, Razi University, Kermanshah, Iran}
\address[edinburgh]{School of Engineering, University of Edinburgh, UK}

\begin{abstract}
Low-resolution quantization is essential to reduce implementation cost and power consumption in massive multiple-input multiple-output (MIMO) systems for 5G and 6G. 
While most existing studies assume perfect channel state information (CSI), we model the impact of coarse quantization noise on both channel estimation and data transmission, yielding a more realistic assessment of system performance under imperfect CSI conditions in the uplink. 
We develop a tight approximation for the bit-error ratio (BER) of uncoded M-QAM with zero-forcing detection, based on the linear minimum mean-square error (LMMSE) channel estimate. 
These analytical results enable compensation strategies that jointly design quantization resolution, transmit power, and pilot length across different numbers of users and base station antennas. We further demonstrate the applicability of the proposed framework through several design scenarios that highlight its effectiveness in optimizing system parameters and improving energy efficiency under quantization constraints. 
For example, in a 16-QAM system, extending the pilot sequence by 2.5 times and lowering transmit power by 0.5 dB enables a 3-bit quantized system to match the BER of the full-resolution case. 
The proposed framework offers a fast and accurate alternative to Monte Carlo simulations, enabling practical system optimization under realistic quantization constraints.
\end{abstract}


\end{frontmatter}

\section{Introduction}
\label{sec:Intro}

\subsection{Motivation and Background} 
Massive MIMO, as a cornerstone of 5G and a foundational element in 6G research,  enables base stations (BSs), equipped with hundreds of antennas, to serve multiple users simultaneously, regardless of whether those users have single or multiple antennas \cite{wang2024tutorial,bjornson2019massive}.
However, having a large number of antennas and radio frequency (RF) chains in massive MIMO results in increased cost, energy consumption, and hardware complexity.  
A common solution is to use low-resolution analog-to-digital converters (ADCs), which consume significantly less power and have simpler designs that reduce hardware cost and complexity \cite{Zhang2025,Gao2024,Zhang2024,AntwiBoasiako2022,Xiang2023}.
They also produce less data, which eases the load on interconnects and digital processing, and makes massive MIMO systems more scalable and efficient.
Nevertheless, using low-resolution ADCs, particularly 1-bit, can degrade system performance due to increased quantization noise. This raises a key question: \textit{Can we maintain performance while benefiting from low-resolution quantization?} 

Achieving a balance between reduced quantization resolution and acceptable performance loss remains one of the main challenges in using low-resolution ADCs.

\subsection{Related Works}
Numerous studies have explored the performance limitations introduced by low-resolution quantization from various perspectives, considering different performance metrics and system scenarios.
Several studies have investigated the capacity of massive MIMO systems with low-resolution quantizers in the uplink mode for maximum ratio combining (MRC), zero-forcing (ZF), and minimum mean-square error (MMSE) detectors \cite{fan2015uplink, Qiao2016, dong2017spectral,yang2022}. These works, using the additive quantization noise model (AQNM) and assuming perfect channel state information (CSI), have provided approximations of the achievable rate under Rayleigh fading channels. 

The impact of reduced quantization resolution on channel estimation has been the focus of several studies. In \cite{jacobsson2015one}, the authors considered systems with 1-bit ADCs, employing a least-squares (LS) channel estimator and analyzing the achievable rate using maximum ratio combining (MRC) detectors. 
In a similar context, \cite{fan2016optimal} used an MMSE estimator to optimize the pilot length for improving the spectral efficiency. Based on Bussgang’s theorem, \cite{li2017channel} and \cite{jacobsson2017throughput} reformulated the linear MMSE (LMMSE) estimator for 1-bit and multi-bit resolutions, respectively, and derived achievable rates for MRC and ZF detectors.  
Additional research has focused on Rician channels with imperfect CSI, using the AQNM model to assess achievable rates for ZF \cite{Ghacham2019} and MRC \cite{Liu2021} detection schemes.
 
In \cite{mollen2016uplink}, the uplink achievable rate of wideband massive MIMO systems using orthogonal frequency-division multiplexing (OFDM) with 1-bit ADCs was investigated, taking channel estimation errors into account. 
To reduce power consumption and enhance achievable rates, mixed-architectures (combining low- and high-resolution ADCs) have been proposed in \cite{zhang2017performance} and \cite{pirzadeh2018spectral}. These works present approximations of the achievable rate for such architectures in the uplink, using the AQNM model for Rician channels and Bussgang’s model for Rayleigh channels, respectively.
 
Additionally, for massive MIMO systems with full-duplex and relay configurations, \cite{kong2017full,Rahimian2020,Dong2020,Xiong2021,Ding2022,AntwiBoasiako2022} evaluated achievable rates and energy efficiency under low-resolution quantization, primarily using the AQNM model. However, none of the aforementioned studies have considered the type and order of modulation in their analytical frameworks. Although the study \cite{li2017channel} included modulation effects, it ultimately relied on Monte Carlo simulations to complete its analysis. 
Studies \cite{Xiang2023,Wu2024} have also explored the ergodic capacity while accounting for modulation, although they still relied on the assumption of perfect CSI.

Most existing studies have focused on analyzing system capacity, while bit error ratio (BER) analysis and the impact of modulation type have received less attention. However, BER is a fundamental metric for assessing the reliability and quality of service (QoS) in practical communication systems. Furthermore, in massive MIMO systems and the 3GPP standard, high-order modulation schemes such as 16-QAM, 64-QAM, and 256-QAM are used \cite{3gpp38.211}.
Coarse quantization introduces significant distortion in such high-order modulation formats.
As a result, analyzing the BER in such scenarios becomes increasingly important and serves as a complement to capacity-focused studies.

Although some studies have addressed error performance metrics, many have focused on analyzing the Symbol Error Ratio (SER) under perfect CSI conditions \cite{Choi2015,saxena2017analysis,Alevizos2018,Ma2019,Zhang2020}. 
However, SER is less applicable in practical systems than BER, as it lacks bit-level resolution.  
The number of studies that specifically investigate BER remains limited, and a substantial portion are based solely on simulation results \cite{zhang2015mpdq,desset2015validation,Temiz2019}.

Existing analytical studies are also confined to specific cases; for example, BER has been analyzed only for QPSK modulation in the downlink mode under perfect CSI \cite{jacobsson2017quantized, Jacobsson2019}, or without considering the effect of quantization noise during channel estimation \cite{Guerreiro2017}.
A study in \cite{Wang2025} examines the BER performance of M-QAM modulation schemes in a uniform circular array-based orbital angular momentum (UCA-OAM) transmission model under perfect CSI, using the AQNM model. It performs an analytical assessment up to a certain point and completes the remaining analysis through simulations.

In our previous work, we proposed an analytic upper bound for the BER performance, which was derived under the assumption of perfect CSI \cite{azizzadeh2019beranalysis}. However, in practical scenarios, quantization noise not only degrades user data but also introduces errors in channel estimation, further reducing system performance compared to the perfect CSI case.

\subsection{Contributions}
%
In this study, we develop an analytical framework that jointly characterizes the effects of coarse quantization on both the channel estimation and data transmission phases of uplink massive MIMO. In contrast to the majority of existing analyses, including our earlier study \cite{azizzadeh2019beranalysis}, which assumed perfect CSI, the proposed framework yields the BER of low-resolution systems under imperfect CSI, where quantization noise corrupts the pilot observations as well as the data. Based on the derived expressions, we analyze the influence of the key system parameters, examine the resulting trade-offs, and propose compensation strategies.

The proposed analytical expressions enable \textit{fast and accurate prediction of system performance}, serving as an efficient alternative to time-consuming and computationally intensive Monte Carlo simulations. They facilitate optimal parameter selection to achieve a target BER, particularly in scenarios where certain parameters are constrained. In addition, they can be integrated into adaptive architectures \cite{Alnajjar2015,Zhou2016}, combined with machine learning or AI algorithms for improved decision making \cite{Li2019,Mashhadi2021}, or even used to generate training data for such models.

We note that Bussgang-based LMMSE (BLMMSE) channel estimators have been derived for 1-bit and multi-bit quantized systems \cite{li2017channel, jacobsson2017throughput}, and related estimators exist for mixed-ADC, relaying, and spatially correlated settings \cite{pirzadeh2018spectral, kong2017full, Dong2020}.
However, these estimators are formulated under their respective quantization models and system structures and are therefore not directly applicable to the AQNM-based framework considered in this work. Accordingly, we derive the LMMSE channel estimator under AQNM that is consistent with the adopted system model. This derivation is essential for incorporating channel estimation error into the subsequent performance analysis and provides the foundation for the contributions that follow.
\begin{itemize}

    \item We develop an analytical framework that captures the effects of coarse quantization on both channel estimation and data transmission under imperfect CSI. Within this framework, we obtain closed-form expressions for the LMMSE channel estimate and the signal-to-interference-plus-quantization-and-noise ratio (SIQNR), and derive a tight BER approximation for uncoded M-QAM with zero-forcing detection.

    \item We show that the derived expressions enable joint design of key system parameters, such as pilot length, transmit power, and quantization resolution, to mitigate quantization-induced performance degradation and achieve BER levels comparable to full-precision systems. Moreover, we provide a closed-form condition in (\ref{eq:tau_q_p_uq}), that specifies how these parameters should be adjusted to achieve full-precision performance.
    
    \item We systematically analyze and validate several practical design scenarios, including balancing parameters to match unquantized BER levels, compensating for performance loss as user count increases, selecting power-efficient quantization resolutions, determining the minimum number of antennas for a target user load and BER, and scaling users under quality-of-service constraints.
    
\end{itemize}

\subsection{Paper Outline}
The remainder of this paper is structured as follows. 
Section \ref{sec:model} describes the system model and a linear quantization model to investigate the effects of coarse quantization on both the channel estimation and the data detection stages. 
In Section \ref{sec:H-estimation}, we derive an LMMSE channel estimate in the presence of quantization error and analyze how it is affected by different parameters (quantization resolution, transmit power, and pilot length). 
Next, we present an SIQNR and a BER expression in Section \ref{sec:sinr_ber_quantized}, taking into account the channel estimation error and quantization error. Monte Carlo simulations are performed in Section \ref{sec:SimulationResults} to validate our proposed analytical expressions. Finally, Section \ref{sec:conclusion} concludes the paper.

\textit{Notations:}
Bold lowercase and uppercase letters denote vectors and matrices, respectively. The superscripts $(\cdot)^T$ and $(\cdot)^H$ represent the transpose and the Hermitian (conjugate transpose). $\mathbb{C}$ denotes the set of complex numbers, and $\mathbb{E}\{\cdot\}$ stands for expectation. $\mathbf{I}_N$ is the $N\times N$ identity matrix, $\mathrm{diag}(\cdot)$ forms a diagonal matrix from its argument, $\mathrm{Tr}(\cdot)$ denotes the trace, and $\|\cdot\|_F$ represents the Frobenius norm.

\section{System Model}
\label{sec:model}
\begin{figure}[t]
		\centering
		\includegraphics[width=1\linewidth]{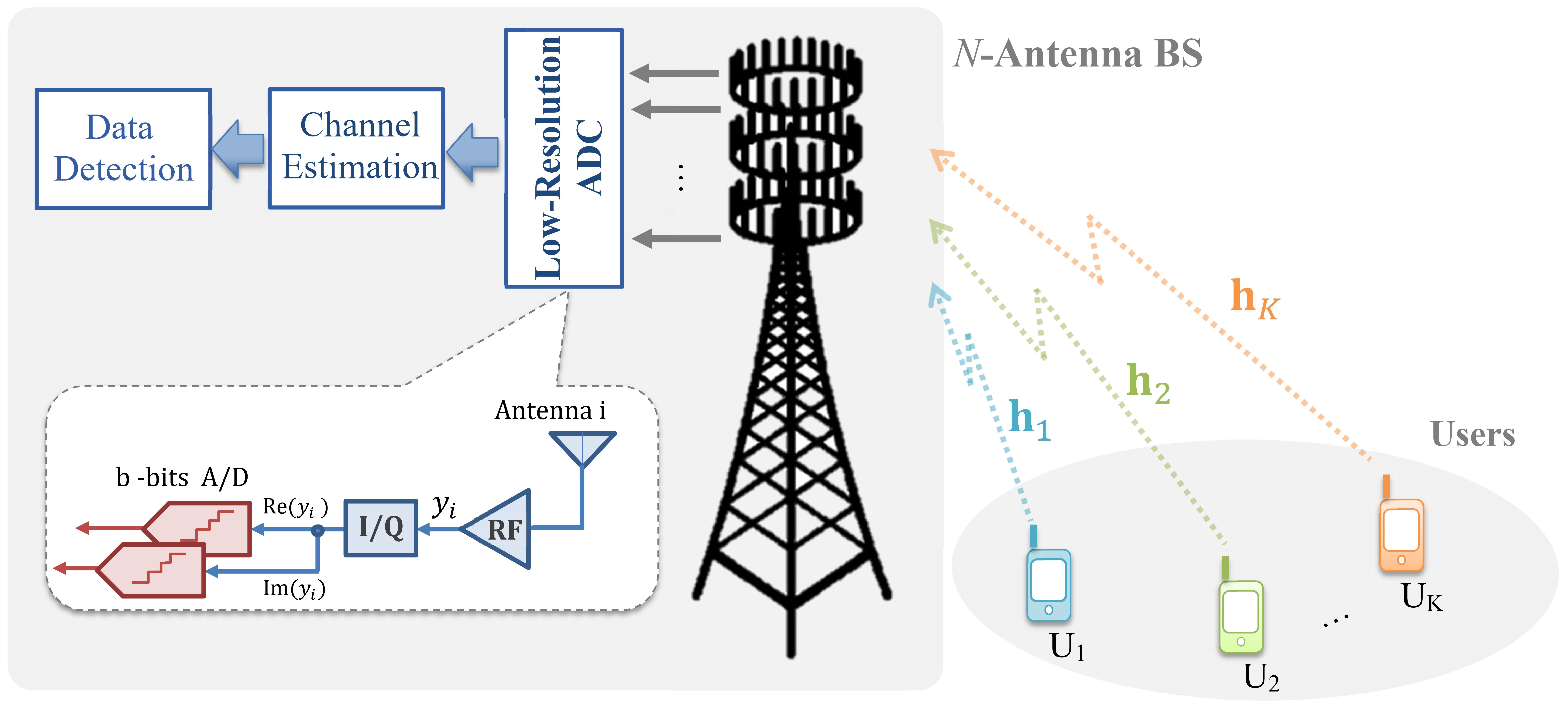}
		\caption{An uplink quantized massive MIMO system with $K$ single-antenna users and one BS having $N$ antennas. }
		\label{fig:ADC_massive}
	\end{figure}

We consider a single cell uplink massive MIMO system with $K$ single-antenna users transmitting their signals to the BS equipped with $N$-antennas. This is illustrated in Fig. \ref{fig:ADC_massive}.
We assume the users transmit a symbol vector $\mathbf s \in \mathbb{C}^{K\times 1}$ where each symbol $s_k$ has a constellation size $M$ and $E\{|s_k|^2\}=1$. Assuming a flat-fading (single-carrier in OFDM) channel model, the received symbol vector $\mathbf y \in \mathbb{C}^{N\times 1}$ at the BS, can be written as
\begin{equation}
\mathbf y =\sqrt{p_u}\mathbf  H \mathbf s +\mathbf n=\sqrt{p_u}\sum_{k=1}^K \mathbf h_k s_k+\mathbf n
\label{eq:y_puHs+n}
\end{equation}
where $\mathbf H\overset{\Delta}{=} [\mathbf h_1,\mathbf h_2,...,\mathbf h_K]\in \mathbb{C}^{N\times K}$ denotes the channel matrix, and  $\mathbf h_k \in \mathbb{C}^{N\times 1}$ is the channel vector between the $k$th user and the BS. The transmit vector of the users is denoted by $\mathbf s=[s_1,s_2,\dots,s_K]^T$, and $\mathbf n \in \mathbb{C}^{N\times 1}$  is the additive white Gaussian noise vector.  The entries of $\mathbf n$ are assumed to be i.i.d Gaussian random variables with zero-mean, unit variance, and are statistically independent of $\mathbf H$. The average transmit power of all users is assumed to be equal and is denoted by $p_u$, which also represents the average signal-to-noise ratio (SNR) for each user.
		
	We see from Fig. \ref{fig:ADC_massive} that $2N$ identical $b$-bit resolution quantizers are needed for the real and imaginary parts of the complex received signals for all antennas. The resulting quantized signal vector can be written as
	\begin{equation}
	\mathbf y_q=Q(\mathbf y)=Q\left(\Re(\mathbf y)\right)+jQ(\Im(\mathbf y))
	\label{eq:yq_Q}
	\end{equation}
	where $Q(\cdot)$ is the quantization function. 
The nonlinear nature of the quantization process complicates the analysis of low-resolution systems. To enable a more tractable analysis, the AQNM model is employed in this work, offering sufficient accuracy in low and moderate SNR regimes and facilitating the analysis of quantized systems \cite{mo2015capacity,orhan2015low}.
%
It has been widely used across quantized MIMO settings, including rate and spectral efficiency analysis \cite{fan2015uplink,Qiao2016,dong2017spectral}, Rician fading channels \cite{Ghacham2019,Liu2021}, relaying and full-duplex systems \cite{AntwiBoasiako2022,Dong2020,Xiong2021,Ding2022}, and recent extensions to emerging system architectures \cite{Zhang2025,Gao2024,Zhang2024,Wang2025}.
The AQNM is more accurate for $b\ge 2$, since at $b=1$ the quantization distortion becomes strongly signal-dependent. Our Monte Carlo results in  Section~\ref{sec:SimulationResults}  show close agreement across all considered resolutions, including $b=1$, in the operating regimes of interest.

Assuming identical $b$-bit resolution ADCs for all received signals, we have \cite{fan2015uplink}
\begin{equation}
\mathbf y_q=Q(\mathbf  y)\approx \alpha\mathbf  y + \mathbf  n_q
\label{eq:yq_aqnm}
\end{equation}
where $\alpha$ is the linear gain, and $\mathbf  n_q$ is the Gaussian quantization noise vector, uncorrelated with the input vector $\mathbf  y$. Having non-uniform and low resolution quantizers with bit-resolution of  $b\le 5$, the values of $\alpha$ are shown in Table \ref{tab:rho}. For higher bit-resolution of $b>5$, an approximate of $\alpha = 1- \frac{\pi\sqrt{3}}{2} 2^{-2b}$ can be applied.
For a fixed channel realization $\mathbf H$, the covariance matrix of $\mathbf n_q$ can be written as \cite{fan2015uplink}
\begin{align}
\mathbf R_{\mathbf n_q\mathbf n_q}  
&=\mathbb{E} \left\{\mathbf n_q \mathbf n_q^H|\mathbf H\right\} 
\approx \alpha(1-\alpha) \diag\left(\mathbf R_{\mathbf y\mathbf y}\right)\nonumber\\
&= \alpha(1-\alpha) \diag\left( p_u \mathbf H\mathbf H^H + \mathbf I_N  \right)
\label{eq:R_nq}
\end{align}
		
\begin{table}
    \renewcommand{\arraystretch}{1.5}
    \centering
    \caption{values of $\alpha$ for $b$-bit resolution ADCs}
    \label{tab:rho}
    \begin{tabular}{|c|c|c|c|c|c|}\hline
        $b$	    & 1		 & 	2	  & 	3	& 	4	  & 5	\\ \hline\hline
        $\alpha$& 0.6366 & 0.8825 & 0.96546 & 0.990503& 0.997501\\ \hline
    \end{tabular}
\end{table}

\emph{Scope and assumptions:} Throughout the paper, we consider uncoded BER over Rayleigh flat-fading channels in a single-cell uplink, employing linear zero-forcing (ZF) detection at the BS. These assumptions are made to keep the analysis tractable and to isolate the interplay between quantization resolution, pilot length, and transmit power, which is the main focus of this work. Uncoded BER provides a conservative, modulation-aware performance measure whose trends generally extend to coded systems.
 
\section{Channel Estimation for Coarsely Quantized Massive MIMO Systems}
\label{sec:H-estimation}

In massive MIMO, the BS estimates the channel matrix $\mathbf{H}$ from uplink pilot signals to enable signal detection and downlink precoding. Assuming a Rayleigh block-fading channel (constant over $T$ symbols), the transmission consists of $\tau$ pilot symbols followed by $(T-\tau)$ data symbols. We consider equal transmit power for pilots and data, a practical case when users cannot adjust power between these transmission phases \cite{li2016much}.
    
Let $\mathbf S_p \in \mathbb{C}^{K\times \tau}$ be the pilot matrix for all users where $\mathbf s_{pi} \in \mathbb{C}^{1\times \tau}$ is the pilot vector of $i$th user, and $\mathbb{E}\{|s_p(i,j)|^2\}=1$.
Considering equal pilot length of $\tau$ (number of pilot symbols) for all users and orthogonality between each pair of pilot vectors, we can write $\mathbf S_p\mathbf S_p^H=\tau\mathbf I_K$. Orthogonality of pilots requires $\tau \geq K$, while $\tau < T$ ensures that at least one data symbol remains in the coherence block. Hence the condition $K \leq \tau < T$ holds throughout the paper. During the channel estimation phase, the received signal of BS  is given by
\begin{equation}
\mathbf Y_p =\sqrt{p_u}\mathbf  H \mathbf S_p +\mathbf N_p
\label{eq:Yp}
\end{equation}
where $\mathbf Y_p\in\mathbb{C}^{N\times\tau}$ is the received signal matrix, and $\mathbf N_p\in\mathbb{C}^{N\times\tau}$ is the additive white Gaussian noise matrix. The entries of $\mathbf N_p$ are assumed to be independent and identically distributed (i.i.d) Gaussian random variables with zero-mean and unit-variance, and are also independent of the channel matrix $\mathbf H$.

The LMMSE estimator of the channel matrix $\mathbf H$ from the observation matrix $\mathbf Y_p$ can be expressed as $\hat{\mathbf H}=\mathbf Y_p\mathbf W$, where $\mathbf W\in\mathbb{C}^{\tau\times K}$ denotes the matrix of complex weights chosen to solve the following minimization problem 
\begin{equation}
\hat{\mathbf H} = \underset{\hat{\mathbf H}}{\arg\min}\, \mathbb{E}\!\left\{\|\mathbf H - \hat{\mathbf H}\|_F^2\right\}
= \underset{\mathbf W}{\arg\min}\, \mathbb{E}\!\left\{\|\mathbf H - \mathbf Y_p \mathbf W\|_F^2\right\}
\label{eq:H_LMMSE_eq}
\end{equation}

For the scenario of full quantization resolution, substituting (\ref{eq:Yp}) into (\ref{eq:H_LMMSE_eq}) and solving yields the channel estimate 
based on the LMMSE estimator as below \cite{hampton2013introduction}:
\begin{equation}
\hat{\mathbf H}=\sqrt{p_u}\ \mathbf Y_p \mathbf S^H_p \big(p_u\mathbf S_p\mathbf S^H_p+\mathbf I_K\big)^{-1}
\label{eq:H_LMMSE}
\end{equation}

The channel estimation error is defined as $\mathbf e = \mathbf H-\hat{\mathbf H}$. Assuming a Rayleigh distributed channel ($\mathbf H$) with zero-mean and unit-variance entries, the estimate $\hat{\mathbf H}$ and the error $\mathbf e$ are orthogonal, with both having i.i.d. zero-mean complex Gaussian entries.
It can be shown that \cite{ngo2013massive}
\begin{align}
&\sigma_{\hat{h}}^2 = \frac{\tau p_u}{1+\tau p_u}, &
\sigma_e^2 =\frac{1}{1+\tau p_u} 
\label{eq:sigma_h_sigma_e}
\end{align}
where $\sigma_{\hat{h}}^2$ and $\sigma_e^2$ are the variances of $\hat{h}_{i,k}$ and $e_{i,k}$, respectively, for $i=1,\cdots,N$ and $k=1,\cdots,K$.

In full-precision quantization, the channel is estimated from the non-quantized $N\times\tau$ complex matrix $\mathbf Y_p$,  with the LMMSE estimator considering only the additive white Gaussian noise (AWGN).
However, practical massive MIMO employing low-resolution quantizers, require the use of the quantized signal $\mathbf Y_{pq}$. Using the AQNM linear quantizer model,
\begin{align}
\mathbf Y_{pq}&=Q(\mathbf Y_p)\approx \alpha \mathbf Y_p+\mathbf N_{qp}
=\alpha\sqrt{p_u}\  \mathbf H \mathbf S_p + [\alpha \mathbf N_p +\mathbf N_{qp}] \nonumber\\
&=\alpha\sqrt{p_u}\  \mathbf H \mathbf S_p+\mathbf N_e
\label{eq:Ypq}
\end{align}
where $\mathbf N_{qp}$ is the quantization noise matrix. Note that $\mathbf N_{qp}$ is not a quantized version of $\mathbf N_p$ in (\ref{eq:Yp}). We define $\mathbf N_e$ as the combined quantization and additive noise. One might suggest replacing $\mathbf Y_p$ in (\ref{eq:H_LMMSE}) with $\mathbf Y_{pq}$ to obtain the quantized channel estimate. However, this approach only considers AWGN and ignores quantization noise. Thus, we need to substitute $\mathbf Y_{pq}$ in the optimization problem (\ref{eq:H_LMMSE_eq}) and solve it again.

\begin{lemma}\label{lemma:H_hat_q}
    The LMMSE estimate of the channel $\mathbf H$ from the quantized received signal vector $\mathbf Y_{pq}$ in (\ref{eq:Ypq}), is given by
    \begin{equation}
    \hat{\mathbf H}_{q}=\frac{1}{\alpha}\sqrt{p_u}\ \mathbf Y_{pq} \mathbf S^H_p \big(p_u\mathbf S_p\mathbf S^H_p+L(\alpha,p_u)\mathbf I_K\big)^{-1}
    \label{eq:H_qLMMSE}
    \end{equation}
    where
    \begin{equation}
    L(\alpha,p_u)=1+(\frac{1-\alpha}{\alpha})( K p_u+1)
    \label{eq:L_alpha_pu}
    \end{equation}
\end{lemma}
\begin{proof}
    The proof is given in   \ref{sec:Appendix_A}.
\end{proof}

\begin{lemma}
    Given the orthogonality between the LMMSE channel estimate $\hat{\mathbf H}_{q}$ in (\ref{eq:H_qLMMSE}) and its estimation error $\mathbf e_q=\mathbf H - \hat{\mathbf H}_q$, and assuming that both $\hat{\mathbf H}_{q}$ and $\mathbf e_q$ have i.i.d. zero-mean complex Gaussian entries, it follows that 
    \begin{align}
    &\sigma_{\hat{h}q}^2 = \frac{\tau p_u}{L(\alpha,p_u)+\tau p_u}, &
    \sigma_{eq}^2 =\frac{L(\alpha,p_u)}{L(\alpha,p_u)+\tau p_u} 
    \label{eq:sigma_hq_hat_eq}
    \end{align}
\end{lemma}
\begin{proof}
    These results are proved in   \ref{sec:Appendix_B}.
\end{proof}
	
Building on (\ref{eq:sigma_hq_hat_eq}), we can examine the impact of varying the low-resolution quantization on channel estimation as follows.
	
\subsection{Increasing the quantization resolution}
Returning to Table \ref{tab:rho}, as we increase the bit-resolution of the quantizers (ADCs at the uplink receiver in our model), the value of $\alpha$ approaches 1. 
As a result, $L(\alpha,p_u)$ which is a positive real number, decreases toward 1. 
Consequently, the variances $\sigma_{\hat{h}q}^2$ and $\sigma_{eq}^2$ in (\ref{eq:sigma_hq_hat_eq}) approach their corresponding values of  $\sigma_{\hat{h}}^2$ and $\sigma_{e}^2$ in (\ref{eq:sigma_h_sigma_e}).

\subsection{Increasing the transmit power}
When the transmit power $p_u$ increases toward infinity, in the absence of quantization error as in (\ref{eq:sigma_h_sigma_e}), the estimation error variance $\sigma_e^2$ goes to zero. 
In contrast, with low-resolution quantization as shown in (\ref{eq:sigma_hq_hat_eq}), we find that:
\begin{equation}
\sigma_{eq}^2 \xrightarrow [\quad p_u\to\infty \quad]{} \frac{(1-\alpha)K}{(1-\alpha)K+\alpha\tau}
\label{eq:lim_sigma2_e}
\end{equation}
This asymptotic behavior indicates that in the presence of quantization error, increasing transmit power alone cannot eliminate the total channel estimation error. 
However, this error can be reduced by either increasing the bit-resolution of quantizers ($\alpha\to 1$) or by extending the pilot length $\tau$, or by optimizing both parameters concurrently.

\subsection{Effects of pilot length}
Examining (\ref{eq:sigma_h_sigma_e}) and (\ref{eq:sigma_hq_hat_eq}) reveals that an increase in the pilot sequence length $\tau$ leads to a decrease in both 
$\sigma_{eq}^2$ and $\sigma_{e}^2$, corresponding to the channel estimation error for quantized and non-quantized pilots, respectively. However, this improvement is more significant for the non-quantized case, due to the fact that $\sigma_{e}^2$ in (\ref{eq:sigma_h_sigma_e}) is only a function of $\tau$ and the transmit power $p_u$, while $\sigma_{eq}^2$ in (\ref{eq:sigma_hq_hat_eq}) is affected by more parameters.

One may ask: \textit{What pilot length $\tau_q$ is required for a low-resolution quantized system to achieve the same estimation error as the non-quantized scenario with pilot length $\tau$?}

Starting with a desired target $\sigma_{eq}^2(\tau_q) \le \sigma_e^2(\tau)$, we have
\begin{equation}
\tau_q \geq  \tau \ \left(\frac{p_u}{p_{uq}}\right) \ L(\alpha,p_{uq})
\label{eq:tau_q>tau_sigma_e}
\end{equation}
where $p_u$ and $p_{uq}$ are the transmit power for full-precision and low-resolution quantized scenarios, respectively. The function $L(\cdot,\cdot)$ is defined in (\ref{eq:L_alpha_pu}).  
This expression reveals that it is possible to compensate the performance loss due to low-resolution quantization by jointly characterizing the trade-off between $\tau_q$ and $p_{uq}$, when $K$, $\alpha$, $\tau$, and $p_u$ are known. 
	
We further analyze this finding in the following sections, focusing on the BER performance of quantized massive MIMO systems.

\section{SIQNR and BER of Quantized Massive MIMO with Imperfect CSI}
\label{sec:sinr_ber_quantized}

Using linear detectors, the $K$ symbol streams for the $K$ users are extracted from the received symbol vector $\mathbf{y}$ at the BS by applying a detection matrix $\mathbf{A} \in \mathbb{C}^{N \times K}$ as 
$\hat{\mathbf{s}} = \mathbf{A}^H \mathbf{y}$,
where $\mathbf{A}$ for zero-forcing (ZF) detection is given by 
\mbox{$\mathbf{A} = \mathbf{H} (\mathbf{H}^H \mathbf{H})^{-1}$} \cite{Qiao2016}.

Following \cite{azizzadeh2019beranalysis}, we now consider a low-resolution quantized system, in which the quantized received vector $\mathbf y_q$ from (\ref{eq:yq_aqnm}) is used instead of $\mathbf{y}$ to update the detection matrix. 
Furthermore, in practice, the BS does not have perfect channel state information and an estimate $\hat{\mathbf H}$ or $\hat{\mathbf H}_q$  is substituted for $\mathbf H$ to update the detection matrix. Thus, we can rewrite the detection matrix $\mathbf A$ for a coarsely quantized system as
\begin{equation}
\mathbf A=\frac{1}{\alpha}\hat{\mathbf H}_q (\hat{\mathbf H}_q^H \hat{\mathbf H}_q)^{-1}
\label{eq:Aq_zf}                                                                                  
\end{equation}
In order to obtain the BER of such system, we need to calculate the SIQNR of data streams for each user from the detector output  $\hat{\mathbf s}$. By employing (\ref{eq:Aq_zf}), we obtain
\begin{align}
    \hat{\mathbf s} &=\mathbf A^H \mathbf y_q\approx\mathbf A^H(\alpha\mathbf y+\mathbf n_q)\nonumber\\
    &=\frac{1}{\alpha}(\hat{\mathbf H}_q^H\hat{\mathbf H}_q)^{-1}\hat{\mathbf H}_q^H(\alpha\sqrt{p_u}\mathbf H\mathbf s+\alpha\mathbf n+\mathbf n_q) \nonumber \\
    &=\frac{1}{\alpha}(\hat{\mathbf H}_q^H\hat{\mathbf H}_q)^{-1}\hat{\mathbf H}_q^H(\alpha\sqrt{p_u}(\hat{\mathbf H}_q+\mathbf e_q)\mathbf s+\alpha\mathbf n+\mathbf n_q) \nonumber \\
%
%
    &=\sqrt{p_u}\mathbf s +\left[(\hat{\mathbf H}_q^H\hat{\mathbf H}_q)^{-1}\hat{\mathbf H}_q^H(\sqrt{p_u}\mathbf e_q\mathbf s+\mathbf n+\frac{1}{\alpha}\mathbf n_q)\right]\nonumber \\
    &=\sqrt{p_u}\mathbf s+\mathbf n_e
    \label{eq:s_hat=pu*s+ne} 
\end{align}
We refer to $\mathbf{n}_e$ as the \textit{effective noise}, which accounts for the additive white Gaussian noise $\mathbf{n}$, quantization noise $\mathbf{n}_q$, and channel estimation error $\mathbf{e}_q = \mathbf{H} - \hat{\mathbf{H}}_q$.
To find the SIQNR of each user, we need to calculate the covariance matrix of  $\mathbf n_e$ as shown in (\ref{eq:E_ne}).  

\begin{figure*}
    \begin{align}
    \mathbb{E}\lbrace \mathbf n_e \mathbf n_e^H\rbrace 
    &=(\hat{\mathbf H}_q^H\hat{\mathbf H}_q)^{-1}\hat{\mathbf H}_q^H \left[\mathbb{E}\lbrace p_u\mathbf e_q \mathbf s \mathbf s^H \mathbf e_q^H\rbrace+\mathbb{E}\lbrace \mathbf n \mathbf n^H\rbrace+\frac{1}{\alpha^2}\mathbb{E}\lbrace\mathbf n_q \mathbf n_q^H\rbrace\right]\hat{\mathbf H}_q(\hat{\mathbf H}_q^H\hat{\mathbf H}_q)^{-1}\nonumber\\
    &=(\hat{\mathbf H}_q^H\hat{\mathbf H}_q)^{-1}\hat{\mathbf H}_q^H \left[ K p_u\sigma_{eq}^2 \mathbf I_N+\mathbf I_N+\frac{(1-\alpha)}{\alpha}\diag(p_u\mathbf H \mathbf H^H+\mathbf I_N)\right] \hat{\mathbf H}_q(\hat{\mathbf H}_q^H\hat{\mathbf H}_q)^{-1}  \nonumber\\
    &=[K p_u\sigma_{eq}^2 +1+\frac{(1-\alpha)}{\alpha}(K p_u+1)](\hat{\mathbf H}_q^H\hat{\mathbf H}_q)^{-1}
    =[K p_u\sigma_{eq}^2 +L(\alpha,p_u)](\hat{\mathbf H}_q^H\hat{\mathbf H}_q)^{-1}
    \label{eq:E_ne}
    \end{align}
\end{figure*}

Assuming equal transmission power for all users, we can write the SIQNR of the $k$th user as 
\begin{equation}
\gamma_{q,k}  = \frac{p_u}{[K  p_u\sigma_{eq}^2 +L(\alpha,p_u)]}\frac{1}{[(\hat{\mathbf H}_q^H\hat{\mathbf H}_q)^{-1}]_{kk}}
\label{eq:gamma_qk}
\end{equation}

The channel estimate $\hat{\mathbf H}_q$ can be normalized as $\bar{\mathbf H} = \hat{\mathbf H}_q / \sigma_{\hat{h}q}$, where entries in $\bar{\mathbf H}$ are i.i.d Rayleigh distributed random variables with zero-mean and unit-variances. Substituting this normalized channel into (\ref{eq:gamma_qk}) gives
\begin{align}
\gamma_{q,k}  &= \frac{p_u\sigma_{\hat{h}q}^2}{[K  p_u\sigma_{eq}^2 +L(\alpha,p_u)]}\frac{1}{[(\bar{\mathbf H}^H\bar{\mathbf H})^{-1}]_{kk}}\nonumber\\
&=\gamma_0\ \frac{1}{[(\bar{\mathbf H}^H\bar{\mathbf H})^{-1}]_{kk}}
\label{eq:gamma_qk_H_bar}
\end{align}
where $\sigma_{\hat{h}q}^2$  and $\sigma_{eq}^2$ are given in (\ref{eq:sigma_hq_hat_eq}). Since $\bar{\mathbf H}$ is Rayleigh distributed, then $\chi_d^2=1/[(\bar{\mathbf H}^H\bar{\mathbf H})^{-1}]_{kk}$ is chi-square distributed with $2(N-K+1)$ degrees of freedom \cite{winters1994impact}. 
As we have equal transmit power and equal SIQNR for all users, we drop the index $k$ and denote the SIQNR of each stream as $\gamma_q=\gamma_0\chi_d^2$. Based on (\ref{eq:gamma_qk_H_bar}) and using (\ref{eq:sigma_hq_hat_eq}), we can express $\gamma_0$ as 
\begin{align}
\gamma_0 &= \frac{p_u^2 \tau}{L(\alpha,p_u)[L(\alpha,p_u)+(K+\tau)p_u]} 
\label{eq:gamma_q0}
\end{align}

Similarly, $\gamma_0$ for full-precision systems is obtained by substituting $\mathbf y$, $\hat{\mathbf H}$, and $\mathbf e$ into (\ref{eq:s_hat=pu*s+ne}) and (\ref{eq:E_ne}). It follows that
\begin{equation}
\gamma_0 = \frac{p_u \sigma_{\hat{h}}^2}{K p_u\sigma_e^2 +1}= \frac{p_u^2 \tau}{1+(K+\tau)p_u} 
\label{eq:gamma_0}
\end{equation}
where $\sigma_{\hat{h}}^2$ and $\sigma_e$ are described in (\ref{eq:sigma_h_sigma_e}).
	
	Using the previous results in \cite{azizzadeh2019beranalysis} (for perfect CSI) together with our current findings in (\ref{eq:gamma_q0}) and (\ref{eq:gamma_0}), we summarize the values of $\gamma_0$ in Table \ref{tab:gamma_0} for different scenarios, based on the assumed channel information ($\mathbf H$ or its estimate $\hat{\mathbf H}_q$) and quantization quality (full precision or coarse quantization).
We evaluate the consistency of expressions in Table \ref{tab:gamma_0} and observe the following properties.

\begin{table}
    \renewcommand{\arraystretch}{1.4}
    \centering
    \caption{Values of $\gamma_0$ for different scenarios at uplink.}
    \label{tab:gamma_0}
    \begin{tabular}{|c|c|c|}
        \hline
        & Full precision & low resolution \\ \hline
        Perfect CSI & $p_u$ & $\frac{p_u}{L(\alpha,p_u)}$\\[1.2ex] \hline
        Imperfect CSI& $ \frac{p_u^2 \tau}{1+(K+\tau)p_u}$ &
        $\frac{p_u^2 \tau}{L(\alpha,p_u)[L(\alpha,p_u)+(K+\tau)p_u]}$\\[2ex] \hline
    \end{tabular}
\end{table}

\subsection{Increasing the Quantization Resolution}
As we increase the bit-resolution $b$ for ADCs, the value of $\alpha$ approaches 1, and $L(\alpha,p_u)\to 1$. Thus, both expressions of $\gamma_0$ at the right hand column, for low-resolution quantization with either full channel knowledge $\mathbf H$ or an estimate $\hat{\mathbf H}_q$, converge to their corresponding full-precision formulas.

\subsection{Increasing the Transmit Power}
When the transmit power $p_u$ (and the SNR per user) goes to infinity, the value of $\gamma_0$ in both expressions  for full-precision quantized system (with $\mathbf H$ or $\hat{\mathbf H}$) goes to infinity and therefore the BER approaches zero.
On the other side, considering low resolution quantization, for perfect CSI we have
\begin{equation}
\gamma_0 \xrightarrow [\quad p_u\to\infty \quad]{} \frac{\alpha}{K(1-\alpha)}
\label{eq:lim_gamma_0_H}
\end{equation}
whereas for imperfect CSI (using $\hat{\mathbf H}_q$), it is
\begin{equation}
\gamma_0 \xrightarrow [\quad p_u\to\infty \quad]{} \frac{\alpha}{K(1-\alpha)} \frac{\alpha\tau}{(K+\alpha\tau)}
\label{eq:lim_gamma_0_H_hat}
\end{equation}
If we assume a pilot length $\tau=K$, then the limit of $\gamma_0$ in (\ref{eq:lim_gamma_0_H_hat}) equals $\frac{\alpha}{K(1-\alpha)} \frac{\alpha}{(1+\alpha)}$. 
Table \ref{tab:rho} indicates that
$\frac{\alpha}{1+\alpha} \to \frac{1}{2}$ as $\alpha \to 1$ with increasing bit resolution.
As a result, we see that even with $p_u\to\infty$, the SINR of the case with imperfect CSI (channel estimate $\hat{\mathbf H}_q$) is about half of its value for perfect CSI. 
Therefore, we need to increase the pilot length as well, to compensate the effects of low-resolution quantization.

\subsection{Increasing the Pilot Length}
Increasing the pilot length $\tau$ toward infinity, both expressions of $\gamma_0$ for imperfect CSI ($\hat{\mathbf H}$ or $\hat{\mathbf H}_q$) in the last row of Table \ref{tab:gamma_0}, would be equal to their corresponding expressions with full CSI in the row above. The reason for such behavior is that the estimation error ($\sigma_e^2$ or $\sigma_{eq}^2$) for both cases of having $\hat{\mathbf H}$ or $\hat{\mathbf H}_q$ goes to zero when $\tau\to\infty$. 
However, as shown in (\ref{eq:tau_q>tau_sigma_e}), a larger pilot length $\tau_q$ is required for a low-resolution quantized system, to achieve the same performance as the full-precision quantized system with the pilot length $\tau$. 
As we have seen from previous discussions, the SINR loss of the coarsely-quantized system can be compensated by increasing either the pilot length or the transmit power of the users. Now we will explore this issue in more detail.

We denote the pair of the transmit power and the pilot length for the full-precision and low-resolution quantized systems by $(p_u,\tau)$ and $(p_{uq},\tau_q)$, respectively. Returning to Table \ref{tab:gamma_0}, to achieve the same $\gamma_0$ for both coarse quantization and full precision, we have
\begin{equation}
\tau_q \geq  \frac{p_u^2\tau L(\alpha,p_{uq})[L(\alpha,p_{uq})+K p_{uq}]}{p_{uq}^2[1+(K+\tau) p_u]-p_{uq}p_u^2\tau L(\alpha,p_{uq})}
\label{eq:tau_q_p_uq}
\end{equation}
Any pair of transmit power and pilot length values $(p_{uq},\tau_q)$ that satisfies (\ref{eq:tau_q_p_uq}), can guarantee an equal or larger SINR for the coarsely quantized system in comparison to the full-precision quantized system with power/pilot settings $(p_u,\tau)$. However, the denominator of inequality in (\ref{eq:tau_q_p_uq}) may be negative for some cases which means we have no solution for $(p_{uq},\tau_q)$ to compensate the SINR loss of low-resolution quantization.

To evaluate the BER performance of the quantized MIMO system, we start from the BER of uncoded M-ary QAM modulation over an AWGN channel in a single-input single-output (SISO) system, which is given by \cite{cho2002general}:
 \begin{equation}
  P_b(\gamma)=\!\!\sum_{k=1}^{\log_2\!\sqrt{M}}\sum_{i=0}^{(1-2^{-k})\sqrt{M}-1}\! 
  F(k,i)\,Q\!\left(\sqrt{\tfrac{3(2i+1)^2\gamma}{M-1}}\right)
  \label{eq:PbAWGN}
  \end{equation}
where $\gamma$ is the SNR for the AWGN channel, and the coefficient $F(k,i)$ is defined as:
\begin{align}
F(k,i) = \frac{2 (-1)^{\lfloor\frac{i.2^{k-1}}{\sqrt{M}}\rfloor} \Bigl(2^{k-1}- \lfloor \frac{i.2^{k-1}}{\sqrt{M}}+\frac{1}{2}\rfloor \Bigr)}{\sqrt{M}\log_2\sqrt{M}}
\end{align}
where $\lfloor\cdot\rfloor$ is the floor function.
Then, BER of a quantized MIMO system can be derived from (\ref{eq:PbAWGN}) by replacing $\gamma$ with $\gamma_q$. Since $\gamma_q=\gamma_0\chi_d^2$ is a function of the random variable $\chi_d^2$, the average BER over a Rayleigh fading channel is calculated by taking the expectation of $P_{b}$ over $\chi_d^2$ as follows:
\begin{align}
\mathrm{BER}=\mathbb{E}_{\chi_{d}^{2}} \big\lbrace  P_{b} (\gamma_0 x)  \big\rbrace =\int P_{b} (\gamma_0 x)f_{\chi_{d}^{2}}(x) dx
\label{eq:berMQAM1}
\end{align}
where $f_{\chi_{d}^{2}}(x)$ is the probability density function of $\chi_d^2$ with $d=2(N-K+1)$ degrees of freedom.
By substituting (\ref{eq:PbAWGN}) into (\ref{eq:berMQAM1}) and solving it (see Appendix D), a tight BER approximation of the quantized MIMO system, which accounts for the effects of low-resolution quantization and channel estimation error, can be obtained as follows:
 \begin{align}
    \mathrm{BER}= \sum_{k=1}^{\log_2\sqrt{M}}\ \sum_{i=0}^{(1-2^{-k})\sqrt{M}-1}F(k,i)B(\mu_i)
\label{eq:berMQAM3}
\end{align}
where $B(\mu_i)$ is:
\begin{equation}
B(\mu_i)=\!\left(\tfrac{1-\mu_i}{2}\right)^{N-K+1}
\sum_{j=0}^{N-K}\!\binom{N-K+j}{j}
\left(\tfrac{1+\mu_i}{2}\right)^{j}
\label{eq:Bmu}
\end{equation}
and using $\gamma_0$ from (\ref{eq:gamma_q0}) for the quantized system under imperfect CSI, the expression for $\mu_i$ can be rewritten as
\begin{equation}
\mu_i=\sqrt{\tfrac{3(2i+1)^2p_u^2\tau}{2(M-1)L(\alpha,p_u)[L(\alpha,p_u)+(K+\tau)p_u]+3(2i+1)^2p_u^2\tau}}
\label{eq:mui}
\end{equation}
Since the $B(\mu_i)$ values are rapidly decreasing for $i>1$, we can further simplify (\ref{eq:berMQAM3}) by keeping only two dominant terms at $i=0,1$:
    \begin{equation}
    \mathrm{BER} \cong \frac{2(\sqrt{M}-1)}{\sqrt{M}\log_2\sqrt{M}}B(\mu_0)+\frac{2(\sqrt{M}-2)}{\sqrt{M}\log_2\sqrt{M}}B(\mu_1)
    \label{eq:BER_MQAM_simple}
    \end{equation}
We explore the accuracy of such an approximation for BER performance in the following section by performing numerical analysis.

\section{Simulation Results}
\label{sec:SimulationResults}
In this section, we validate the proposed analytical expressions for the channel estimation error and BER, as derived in (\ref{eq:sigma_hq_hat_eq}) and (\ref{eq:BER_MQAM_simple}), using Monte Carlo simulations. Their accuracy is assessed considering different quantization resolutions, modulation orders, and system configurations. The results also show how the analytical framework enables the design of compensation strategies to mitigate the impact of coarse quantization. 

Additionally, we evaluate several practical scenarios to demonstrate how our proposed expressions can guide the selection of system parameters in low-resolution massive MIMO deployments. These results show that performance levels comparable to full-precision systems can be achieved while meeting power and BER constraints.

For each channel realization, a long sequence ($10^5$ to $10^6$ symbols) of equi-probable M-QAM symbols is generated per user. The received vector $\mathbf{y}$ is computed using (\ref{eq:y_puHs+n}) and quantized at each antenna using the non-linear quantizers described in \cite{max1960quantizing,azizzadeh2019beranalysis}.

\subsection{Channel Estimation}
Equation~(\ref{eq:tau_q>tau_sigma_e}) demonstrates that, by properly tuning \( p_{uq} \) and \( \tau_q \), the channel estimation error in a low-resolution quantized system can be made virtually indistinguishable from that of its unquantized equivalent. 
 
Table~\ref{tab:tau_q} illustrates this by showing the required values of $\tau_q$ for $p_{uq}=0$ and 3 dB, and for $K=10$ and 20 users, assuming $\tau$ and $p_u=0$ dB for the unquantized case. For instance, with $K = 20$, $p_u = 0$ dB, and 3-bit resolution ADCs ($b = 3$), the condition $\tau_q\ge 1.75\tau$ must be satisfied. This implies that the pilot sequence length in a 3-bit quantized system needs to be approximately 1.75 times longer to match the performance of the full-precision system. However, by increasing the transmit power by 3 dB, the requirement is relaxed to $\tau_q \ge 1.23\tau$.
Equation (\ref{eq:tau_q>tau_sigma_e}) can sometimes yield $\tau_q < K$ at high resolutions and powers. To guarantee orthogonal pilots, this value is rounded up to $\tau_{q,\min} = K$ in practice.

\begin{figure}[t]
    \centering
    \begin{subfigure}{0.47\linewidth}
        \includegraphics[width=\linewidth]{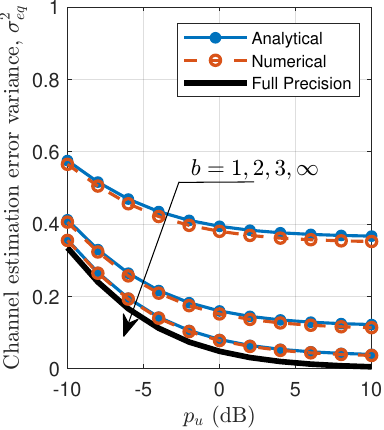}
       \caption{}
        \label{fig:sigma2_eq_tau_K}
    \end{subfigure}
    \begin{subfigure}{0.46\linewidth}
        \includegraphics[width=\linewidth]{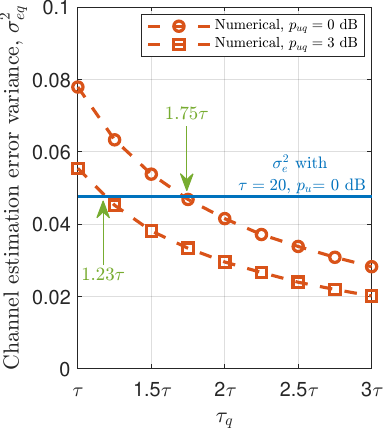}
        \caption{}
        \label{fig:sigma2_eq_tau_2K}
    \end{subfigure}
    \caption{Channel estimation error variance $\sigma_{eq}^2$ versus (a) $p_u$ given $\tau=\tau_q=K$, and (b) $\tau_q$ given $K=20$ and $b=3$ }
    \label{fig:sigma2_eq}
\end{figure}

Figure~\ref{fig:sigma2_eq} presents the behavior of the channel estimation error variance $\sigma_{eq}^2$ as a function of the transmit power $p_{uq}$ in (a) and the pilot length $\tau_q$ in (b). The analytical expressions in (\ref{eq:sigma_hq_hat_eq}) closely align with Monte Carlo simulations assuming $K = 20$.

As a result of the above discussions, \textit{adjusting both the transmit power and the pilot length can mitigate the performance loss in channel estimation for low-resolution quantized systems.}

\begin{table}
    \renewcommand{\arraystretch}{1.5}
    \centering 
    \caption{Required values of \(\tau_q\) from (\ref{eq:tau_q>tau_sigma_e}) for a quantized system using \(b\)-bit ADCs, to match the channel estimation error of the unquantized case with a given \(\tau\) pilots, \(K=10, 20\) users and \(p_u=0\,\mathrm{dB}\).}
    \label{tab:tau_q}
    \begin{tabular}{|c||c|c|c|c|c|}\hline
       & $b$   & 1		 & 	2	  & 	3	& 	4	  \\ \hline\hline
        $K=10$, $p_{uq}=0$ dB& $\tau_q\ge$	& $7.28\tau$ & $2.46\tau$ & $1.39\tau$ & $1.11\tau$\\ \hline
        $K=10$, $p_{uq}=3$ dB& $\tau_q\ge$	& $6.49\tau$ & $1.90\tau$ & $0.88\tau$ & $0.60\tau$\\ \hline
        $K=20$, $p_{uq}=0$ dB& $\tau_q\ge$	& $12.99\tau$ & $3.80\tau$ & $1.75\tau$ & $1.20\tau$\\ \hline       
       $K=20$, $p_{uq}=3$ dB& $\tau_q\ge$	& $12.20\tau$ & $3.23\tau$ & $1.23\tau$ & $0.70\tau$\\ \hline
    \end{tabular}
\end{table}

\subsection{BER Performance}
We numerically validate the BER approximation in (\ref{eq:BER_MQAM_simple}) for uplink M-QAM massive MIMO systems with imperfect CSI due to channel estimation errors and low-resolution quantization. This validation is performed through Monte Carlo simulations for 16-QAM, 64-QAM, and 256-QAM modulation schemes. Several practical scenarios are also presented to demonstrate the applicability of the analytical results.

We assume equal transmit power $p_u$ for both pilot and data sequences for all users. \textit{Note that $p_u$ represents the SNR per user, while $E_b/N_0$ denotes the SNR per user per bit}. The relation between $p_u$ and $E_b/N_0$ is given by:
\begin{equation}
    p_u = \frac{\sigma^2_s}{\sigma^2_n} = \frac{E_s}{N_0/2} = \frac{E_b\log_2 M}{N_0/2}
    \label{eq:pu} 
\end{equation}
where $\sigma^2_s$ is the signal (or symbol) power,  $\sigma^2_n= N_0/2$ is the noise power, and $N_0$ is the noise power spectral density. $E_s$ and $E_b$ are symbol and bit energies, respectively, with $E_s = E_b\log_2 M$ for M-QAM modulations \cite{proakis2007digital}, and $M$ is the modulation order. 
From this point onward, we use $E_b/N_0$ instead of $p_{uq}$ and $p_u$, as defined in (\ref{eq:pu}), to facilitate performance comparison across different M-QAM modulation schemes.

\begin{figure}[t]
    \centering
    \begin{subfigure}{0.49\linewidth}
        \includegraphics[width=\linewidth]{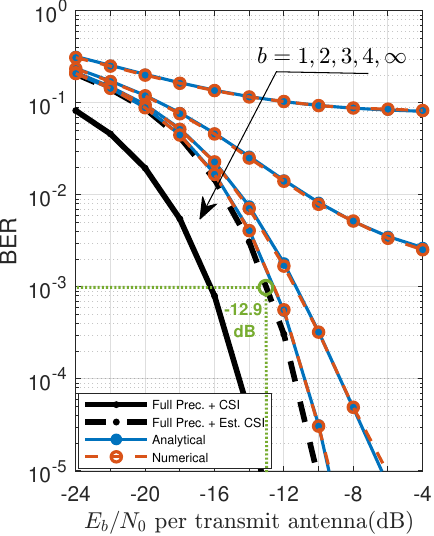}
        \caption{$\tau=K$, 16-QAM}
        \label{fig:BER_16QAM_tau_K}
    \end{subfigure}
    \begin{subfigure}{0.49\linewidth}
        \includegraphics[width=\linewidth]{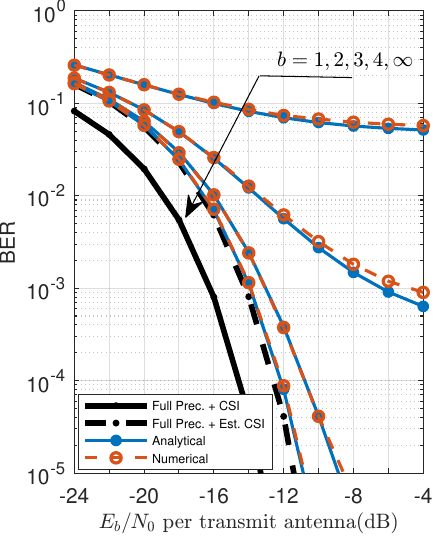}
        \caption{$\tau=2K$, 16-QAM}
        \label{fig:BER_16QAM_tau_2K}
    \end{subfigure}
    \caption{BER of uplink massive MIMO with $N\!=\!256$, $K\!=\!20$, 16-QAM, $b$-bit ADCs, imperfect CSI: (a) $\tau\!=\!K$, (b) $\tau\!=\!2K$.}
    \label{fig:BER_16QAM}
\end{figure}

\begin{figure}[t]
    \centering
    \begin{subfigure}{0.49\linewidth}
        \includegraphics[width=\linewidth]{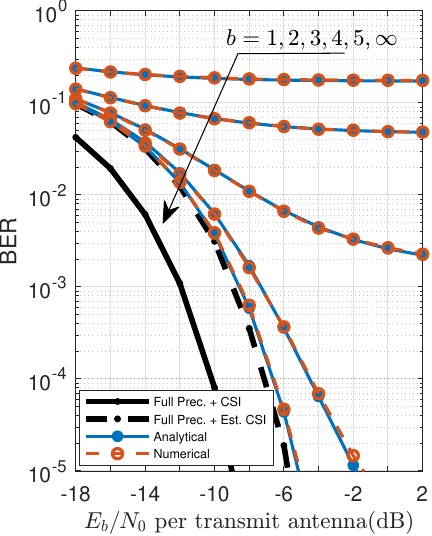}
        \caption{$\tau=K$, 64-QAM }
        \label{fig:BER_64QAM_tau_K}
    \end{subfigure}
    \begin{subfigure}{0.49\linewidth}
        \includegraphics[width=\linewidth]{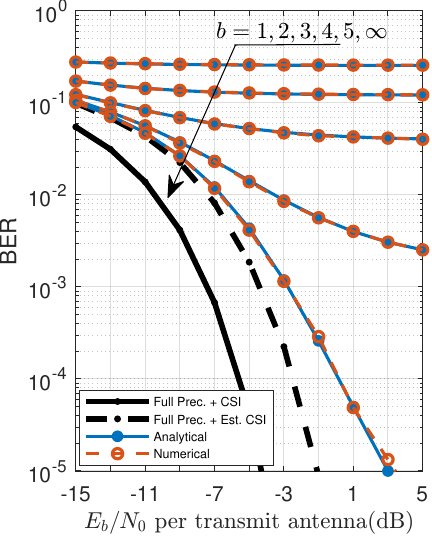}
        \caption{$\tau=K$, 256-QAM}
        \label{fig:BER_256QAM_tau_K}
    \end{subfigure}
    \caption{BER of uplink massive MIMO with $\tau\!=\!K\!=\!20$, $b$-bit ADCs, and imperfect CSI for (a) 64-QAM, (b) 256-QAM.}
    \label{fig:BER_64QAM}
\end{figure}

Figures \ref{fig:BER_16QAM} and \ref{fig:BER_64QAM} illustrate the achievable BER performance of uplink massive MIMO systems with low-resolution ADCs and imperfect CSI (due to channel estimation error), as a function of $E_b/N_0$. Both figures assume $N=256$ antennas for BS and $K=20$ users. In Figure~\ref{fig:BER_16QAM}, uncoded 16-QAM is considered with two different pilot lengths: (a) $\tau=K$ and (b) $\tau=2K$. In contrast, Figure \ref{fig:BER_64QAM} explores higher-order modulations with fixed pilot length $\tau=K$, showing BER performance for (a) 64-QAM and (b) 256-QAM. Numerical results for all modulation schemes and bit resolutions $b$ closely match the analytical results, confirming the validity of the BER approximation given in (\ref{eq:BER_MQAM_simple}).
For comparison, two BER bounds are plotted: (i) an ideal scenario with no quantization or channel estimation error (solid black line, "Full Prec. + CSI", i.e., full precision and perfect channel state information), and (ii) with channel estimation error for a given pilot length $\tau$, but no quantization error (dashed black line, "Full Prec. + Est. CSI"). 
In Figure~\ref{fig:BER_16QAM}(a) and (b), we see that the gap between these two curves decreases as $\tau$ increases from $K$ to $2K$, illustrating the effect of jointly varying $\tau$ and $E_b/N_0$ on the BER performance.

We observe a severe BER degradation when using low-resolution ADCs: 1-bit for 16-QAM, up to 2-bit for 64-QAM, and up to 3-bit for 256-QAM. Under these conditions, achieving a BER of $10^{-2}$ is not possible%
\footnote{This limitation could potentially be mitigated using techniques such as spatial-domain oversampling or time-domain signal repetition. However, these are beyond the scope of this work, which focuses on uncoded M-QAM modulations.}.
Interestingly, 2-bit 16-QAM, 3-bit 64-QAM, and 4-bit 256-QAM show a similar behavior. For squared $M$-QAM modulations with $M=2^{2m}$ (e.g. ${4~(m=1), 16~(m=2), 64~(m=3)}$), achieving reliable performance requires at least $b=m$ quantization bits.

\begin{observation}
The minimum quantization bit resolution required for squared $M$-QAM modulations is proportional to $\log_2{\sqrt M}$, where $M$ is the modulation order.
\end{observation}

Figures \ref{fig:BER_16QAM}-\ref{fig:BER_64QAM} also indicate that, under constant transmit power, both longer pilots and higher quantization resolution improve BER, with quantization having a greater impact since it affects both channel estimation and data detection, whereas pilot length only influences estimation.

\subsection{Performance in Different Scenarios}
In what follows, we explore several practical scenarios to demonstrate the applicability of our proposed expressions in the context of system parameter selection.

\subsection*{Scenario 1: Balancing Parameters in Quantized Systems to Achieve Unquantized BER Levels}
\begin{figure}[t]
    \centering
    \begin{subfigure}{0.47\linewidth}
        \includegraphics[width=\linewidth]{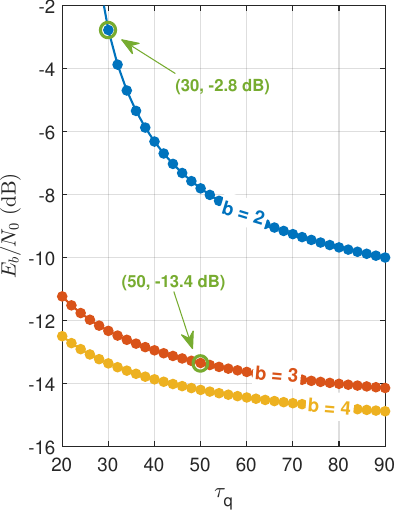}
        \caption{}
        \label{fig:Scenario1_a}
    \end{subfigure}
    \begin{subfigure}{0.49\linewidth}
        \includegraphics[width=\linewidth]{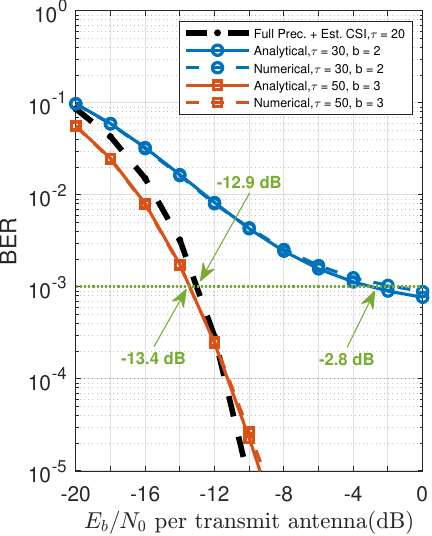}
        \caption{}
        \label{fig:Scenario1_b}
    \end{subfigure}
    \caption{(a) Required $\tau_q$ and $E_b/N_0$ values to achieve a BER of $10^{-3}$ using (\ref{eq:tau_q_p_uq}) for $b$-bit ADCs, $N\!=\!256$, $K\!=\!20$, 16-QAM. (b)~BER variation with $\tau_q$ and $E_b/N_0$ for the same system. }
    \label{fig:Scenario1}
\end{figure}

We target a BER of $10^{-3}$ (chosen only for simulation; other values can also be used). As shown in Figure~\ref{fig:BER_16QAM_tau_K}, for full-precision quantization with imperfect CSI, the required $E_b/N_0$ for 16-QAM is $-12.9$~dB. 
Using (\ref{eq:tau_q_p_uq}) and the target BER of $10^{-3}$, Figure~\ref{fig:Scenario1_a} illustrates the pilot length $\tau_q$ and $E_b/N_0$ required for systems with low-resolution quantizers. Each point on the curves represents a feasible $(\tau_q, E_b/N_0)$ pair that achieves the target BER, indicating either the minimum pilot length for a given $E_b/N_0$ or the minimum $E_b/N_0$ for a given $\tau_q$.  

For example, Figure~\ref{fig:BER_16QAM_tau_K} shows that with 2-bit 16-QAM, the target BER cannot be achieved when $\tau_q = K$. Hence, both the pilot length and transmit power must be increased. As illustrated in Figure~\ref{fig:Scenario1_a}, achieving the target BER requires an $E_b/N_0$ of $-2.8$~dB with a $\tau_q$ of $1.5K$ (30 symbols) for 2-bit quantization, and $-13.4$~dB with a $\tau_q$ of $2.5K$ (50 symbols) for 3-bit quantization.

Figure~\ref{fig:Scenario1_b} validates these findings by comparing Monte Carlo simulations and analytical results for 2-bit, 3-bit, and full-precision quantization with channel estimation (denoted as "Full Prec.~+~Est.~CSI"). In this case, without quantization error, the target BER is achieved using $\tau \!=\! K$ and an $E_b/N_0$ of $-12.9$~dB. In contrast, 2-bit quantization with $\tau_q = 1.5K$ requires an $E_b/N_0$ of $-2.8$~dB and an additional $10.6$~dB of power, which is impractical. However, 3-bit quantization with a $\tau_q$ of 50 symbols ($2.5K$) achieves the target BER at an $E_b/N_0$ of $-13.4$~dB, $0.5$~dB lower than the full-precision case. Other $(\tau_q, E_b/N_0)$ combinations are also possible; these are just two examples. 
Therefore, using (\ref{eq:tau_q_p_uq}), we can select appropriate $(\tau_q, E_b/N_0)$ pairs to compensate for coarse quantization and match the performance of full-precision systems. 
 
\begin{observation}
BER performance loss of an uplink massive MIMO system due to coarse quantization can be compensated by jointly designing the pilot length and the transmit power.
\end{observation} 

In applications prioritizing battery life over data rate (e.g., IoT sensors), longer pilot lengths can be used to reduce transmit power while maintaining a desired BER. Conversely, high-data-rate applications (e.g., video streaming) can achieve their performance with shorter pilot lengths and adjusted transmit power.

\subsection*{Scenario 2: Compensating for BER Degradation as User Count Scales Up}
\begin{figure}[t]
    \centering
    \begin{subfigure}{0.50\linewidth}
        \includegraphics[width=\linewidth]{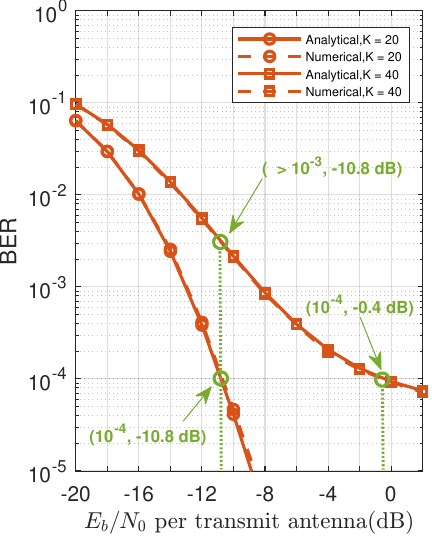}
        \caption{}
        \label{fig:Scenario2_a}
    \end{subfigure}
    \begin{subfigure}{0.48\linewidth}
        \includegraphics[width=\linewidth]{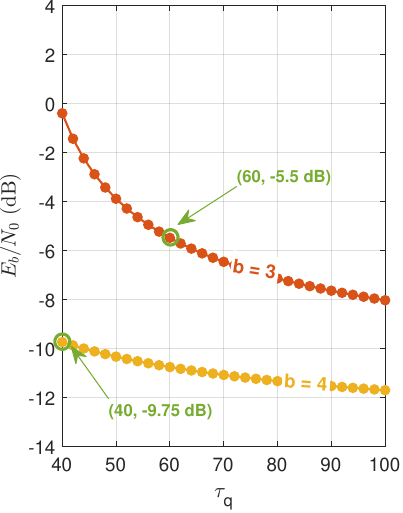}
        \caption{}
        \label{fig:Scenario2_b}
    \end{subfigure}
    \begin{subfigure}{0.49\linewidth}
        \includegraphics[width=\linewidth]{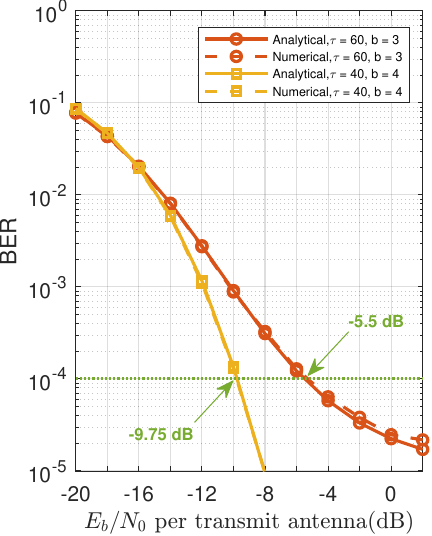}
        \caption{}
        \label{fig:Scenario2_c}
    \end{subfigure} 
    \caption{BER analysis with $N = 256$ and 16-QAM modulation: (a) BER vs. $E_b/N_0$ with 3-bit ADCs, $K = 20, 40$ and $\tau = 40$; (b) Required $(\tau_q, E_b/N_0)$ pairs for BER $= 10^{-4}$, $K = 40$; (c) BER vs. $E_b/N_0$ with 3-4 bit ADCs, $K = 40$, $\tau = 40, 60$.}
    \label{fig:Scenario2}
\end{figure}

Consider a 3-bit quantized system employing 16-QAM modulation, a pilot length of $\tau\!=\!40$, serving $K\!=\!20$ users with a target BER of $10^{-4}$. As shown in Figure~\ref{fig:Scenario2_a}, when the number of users $K$ increases from 20 to 40, the error ratio exceeds $10^{-3}$. If one attempts to compensate for this degradation solely by increasing the transmit power, Figure~\ref{fig:Scenario2_a} shows that the required $E_b/N_0$ must rise from $-10.8$~dB to $-0.4$~dB, which is not a practical solution. The objective is to determine how parameters such as resolution, pilot length, and transmit power should be adjusted to maintain the desired quality of service.

Using (\ref{eq:tau_q_p_uq}) and the target BER of $10^{-4}$, Figure~\ref{fig:Scenario2_b} presents the required combinations of pilot length $\tau_q$ and $E_b/N_0$ for systems with 3- and 4-bit quantization, assuming $K\!=\!40$ users. Each point on the curves represents a feasible $(\tau_q, E_b/N_0)$ pair that satisfies the target BER. For example, with $b = 3$, increasing the pilot length from 40 to 60 symbols and setting $E_b/N_0$ to $-5.5$~dB achieves the desired performance. Alternatively, for $b = 4$, the same pilot length of 40 symbols and an $E_b/N_0$ of $-9.75$dB is sufficient to maintain the error ratio. However, this BER target cannot be met with 2-bit resolution, regardless of pilot length or transmit power, unless a lower-order modulation scheme such as QPSK is employed. Figure~\ref{fig:Scenario2_c} provides Monte Carlo simulation results that validate the analytical predictions.

\subsection*{Scenario 3: Power-efficient ADC resolution at target BER}
\begin{figure}[t]
    \centering
    \begin{subfigure}{0.47\linewidth}
        \includegraphics[width=\linewidth]{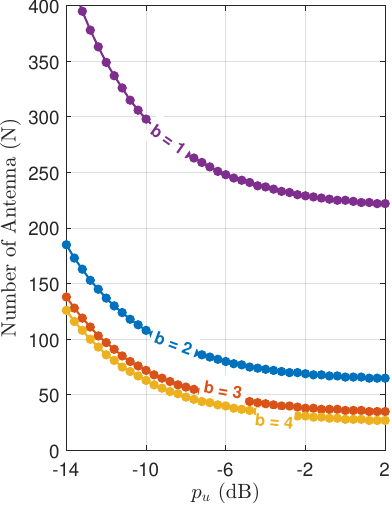}
        \caption{}
        \label{fig:Scenario3_a}
    \end{subfigure}
    \begin{subfigure}{0.48\linewidth}
        \includegraphics[width=\linewidth]{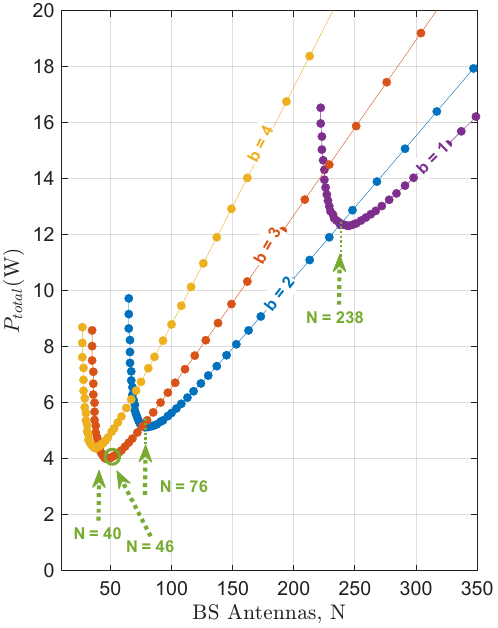}
        \caption{}
        \label{fig:Scenario3_b}
    \end{subfigure}
    \caption{(a) Required $(N, p_u)$ pairs to achieve BER $\!=\!10^{-3}$ for QPSK, $K\!=\!20$, and $\tau \!=\! 40$, under 1–4 bit quantization. (b)~Corresponding total power consumption $P_{\text{total}}$ versus $N$ using values from (a).}
    \label{fig:Scenario3}
\end{figure}

Although low-resolution ADCs consume less power, they degrade BER performance due to increased quantization noise. To meet a given quality-of-service (QoS) requirement, higher transmit power or a larger number of antennas is needed. In contrast, high-resolution ADCs consume more power but can achieve the same QoS with lower transmit power or fewer antennas. The goal is to find the \textit{optimal ADC resolution and number of BS antennas} that minimize total power consumption, including both transmit and ADC processing power, while meeting a QoS constraint (e.g., BER = $10^{-3}$). 
To this end, the set of $(p_u, N)$ pairs that satisfy the target BER for each ADC resolution is derived using equation~(\ref{eq:BER_MQAM_simple}). For example, Figure~\ref{fig:Scenario3_a} shows these pairs for $K\!=\!20$ users, pilot length $\tau\!=\!40$, and QPSK modulation. Then, the total power consumption is given by
\begin{equation}
P_{\text{total}} = K P_{tx} + N \cdot (2P_{\text{ADC}} + P_{\text{RF}}) + P_{\text{Other}},
\end{equation}
where $P_{tx}$ is the radiated transmit power of each user,  $P_{\text{RF}}$ is the RF front-end power per antenna, $P_{\text{Other}}$ is the power of other system components.
Since $p_u$ represents the normalized transmit SNR in the signal model, the corresponding physical transmit power is given by $P_{\mathrm{tx}} = p_u \sigma_n^2/\beta_{\mathrm{ref}}$, 
where $p_u$ is expressed in linear scale, $\sigma_n^2$ is the total receiver noise power, and $\beta_{\mathrm{ref}}$ denotes the reference large-scale channel gain between each user and the BS. The ADC power consumption is given by~\cite{Liu2019}
\begin{equation}
P_{\text{ADC}} = \text{FOM} \cdot f_s \cdot 2^b,
\label{eq:P_ADC}
\end{equation}
where the figure of merit (FOM) represents the energy consumed per conversion step, expressed in fJ/conversion step,  and $f_s$ representing the sampling frequency. The values of $P_{\text{RF}}$ and $P_{\text{Other}}$ are assumed to be constant.

For simulation purposes, a transmission bandwidth of $400~\mathrm{MHz}$ is considered to evaluate the impact of high sampling rates on ADC power consumption in wideband massive MIMO systems. Accordingly, the total receiver noise power is calculated as
\[
\sigma_n^2~[\mathrm{dBm}] = N_0 + 10\log_{10}(B) + \mathrm{NF},
\]
where $N_0=-174~\mathrm{dBm/Hz}$ is the thermal noise power spectral density and $\mathrm{NF}=5~\mathrm{dB}$ is the receiver noise figure, yielding $\sigma_n^2 \approx -83~\mathrm{dBm}$ for $B=400~\mathrm{MHz}$. The sampling rate of each in-phase and quadrature ADC is set to $f_s=1~\mathrm{GS/s}$. The ADC model employs $\mathrm{FOM}=1432.1~\mathrm{fJ/conversion~step}$ \cite{Liu2019}, giving $\mathrm{FOM}\times f_s = 1.4321~\mathrm{mW}$. From (34), the power consumption of a single $b$-bit ADC is $P_{\mathrm{ADC}} = 1.4321\times 2^b~\mathrm{mW}$, yielding $2.8642$, $5.7284$, $11.4568$, and $22.9136~\mathrm{mW}$ for $b=1$, $2$, $3$, and $4$, respectively. 
The RF front-end power per antenna is set to $P_{\mathrm{RF}} = 20$ mW.
A reference large-scale channel gain of $\beta_{\mathrm{ref}}=10^{-10}$, corresponding to a reference path loss of $100~\mathrm{dB}$, is used to map the normalized transmit SNR to the corresponding physical transmit power.

Figure~\ref{fig:Scenario3_b} plots the total power consumption $P_{\text{total}}$ versus the number of antennas $N$ for 1- to 4-bit quantization levels. For each resolution $b$, the corresponding $(p_u, N)$ pair that satisfies the BER constraint is obtained from Figure~\ref{fig:Scenario3_a} and used to compute $P_{\text{total}}$. As observed, the minimum total power occurs at $b = 3$ bits and $N\!=\! 46$ antennas.

Another insight from Figure~\ref{fig:Scenario3_b} is the crossing point between the $P_{\text{total}}$ curves for 1-bit and 2-bit quantization at $N\!=\!238$. For antenna counts below this threshold (with a target BER of $10^{-3}$ and $\tau\!=\!40$), \textit{a system with 1-bit ADCs consumes more power}. This is because the higher transmit power required to meet the BER target with 1-bit resolution outweighs the power savings from using lower-resolution ADCs.
Similarly, a crossing point at $N\!=\!76$ indicates that \textit{for systems with a small number of BS antennas, 3- and 4-bit quantization are more energy-efficient than 1- and 2-bit configurations}.

\subsection*{Scenario 4: Minimum Antennas Needed to Support a Given Number of Users at a Desired BER}

\begin{figure}[t]
    \centering
    \begin{subfigure}{0.48\linewidth}
        \includegraphics[width=\linewidth]{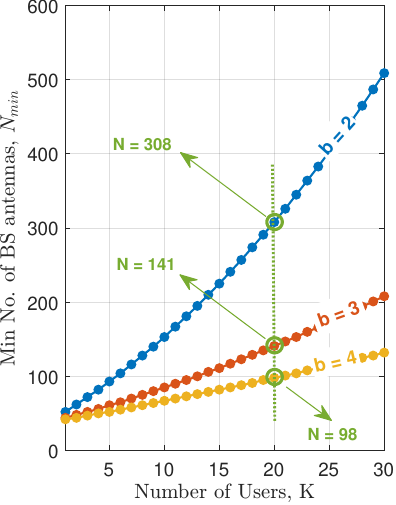}
        \caption{}
        \label{fig:Scenario4_a}
    \end{subfigure}
    \begin{subfigure}{0.492\linewidth}
        \includegraphics[width=\linewidth]{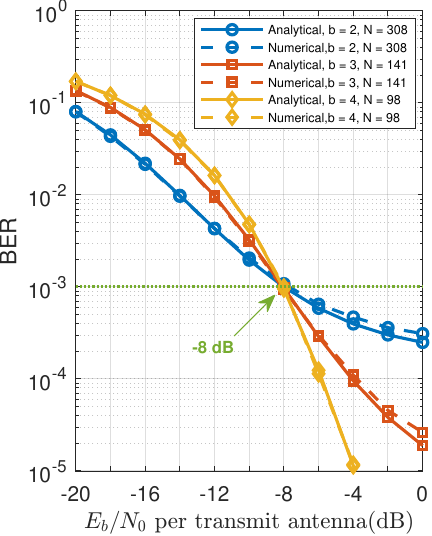}
        \caption{}
        \label{fig:Scenario4_b}
    \end{subfigure}
    \caption{(a) Min. number of BS antennas $N_{\min}$ required to achieve a BER of $10^{-3}$ vs.\ number of users $K$ for 2–4~bit ADCs, assuming 16-QAM and $\tau\!=\!30$. (b)~BER vs.\ $E_b/N_0$ for $K\!=\!20$ using the $N_{\min}$ values from (a).}
    \label{fig:Scenario4}
\end{figure}

In this scenario, we investigate the minimum number of BS antennas ($N_{\min}$) required to support a given number of users $K$ at a target BER of $10^{-3}$. Assuming 16-QAM modulation, a transmit power of $-8$~dB in terms of $E_b/N_0$, and a pilot length of $\tau\!=\!30$, the $N_{\min}$ values for various quantization resolutions are derived using (\ref{eq:BER_MQAM_simple}) and illustrated in Figure~\ref{fig:Scenario4_a}. For instance, when $K\!=\!20$, the required number of antennas for 2-, 3-, and 4-bit resolutions are 308, 141, and 98, respectively. 
Figure~\ref{fig:Scenario4_b} presents the BER curves from both the analytical analysis and Monte Carlo simulations, confirming that at $E_b/N_0 = -8$~dB, the target BER is reliably achieved for the corresponding $N_{\min}$ at each resolution level.

\subsection*{Scenario 5: Max. Number of users supportable at target BER}
\begin{figure}[t]
    \centering
    \includegraphics[width=0.79\linewidth]{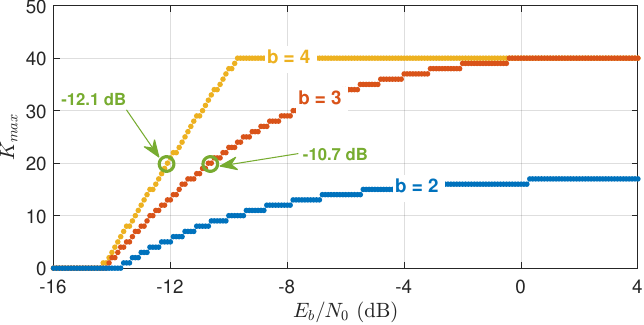}
    \caption{Max. number of supportable users $K_{\max}$ vs.\ $E_b/N_0$ for 2-4~bit quantization, with 16-QAM, $N\!=\!256$, $\tau\!=\!40$, and a target BER of $10^{-4}$. }
\label{fig:Scenario5}
\end{figure}

This scenario aims to determine the maximum number of supported users ($K_{\max}$) for each resolution, ensuring that the system’s BER remains within the acceptable range (i.e., $\text{BER} \leq 10^{-4}$). 
We assume a configuration with $N = 256$ antennas, a pilot length of $\tau = 40$, and 16-QAM modulation. 
By using equation~(\ref{eq:BER_MQAM_simple}), Figure~\ref{fig:Scenario5} presents the derived $K_{\max}$ values for different resolutions under varying transmit power levels. For example, with this antenna setup, up to 20 users can be supported at 3-bit and 4-bit resolutions, corresponding to $E_b/N_0$ values of $-10.7$~dB and $-12.1$~dB, respectively. However, at 2-bit resolution, supporting 20 users with a BER $\leq 10^{-4}$ is not feasible at any transmit power. These findings are consistent with the Monte Carlo simulation results shown in Figure~\ref{fig:BER_16QAM_tau_2K}.

\section{Conclusion}
\label{sec:conclusion}




	
This paper investigated the combined effects of low-resolution ADC quantization and imperfect CSI on the uncoded BER of uplink massive MIMO systems. Based on the AQNM, an LMMSE channel estimator was derived for quantized pilot observations, alongside closed-form expressions for the channel estimation error variance, the post-ZF SIQNR, and a tight BER approximation for uncoded $M$-QAM. Monte Carlo simulations across different modulation orders, ADC resolutions, and system configurations demonstrated close agreement with the analytical results.

The derived expressions characterize the trade-offs among ADC resolution, transmit power, pilot length, user load, and the number of BS antennas. Furthermore, they enable the joint selection of transmit power and pilot length to compensate for quantization-induced degradation. For example, in a 16-QAM system, extending the pilot sequence by a factor of 2.5 allows a 3-bit quantized receiver to match the BER of the full-resolution case at 0.5 dB lower transmit power. Such compensation is feasible provided the resolution satisfies $b \ge \log_2 \sqrt{M}$; below this threshold, no combination of power and pilot length can attain full-precision performance.

The framework further determines the minimum number of BS antennas and the maximum number of supportable users for a target BER, supporting total-power optimization across various ADC resolutions and array sizes. For QPSK at a BER of $10^{-3}$ under the adopted power model, the minimum total power occurs at 3-bit ADCs with $N = 46$ antennas. This demonstrates that using higher ADC resolutions does not necessarily increase total receiver power, as coarser ADCs may demand additional antennas or transmit power to meet the same BER requirement.

Overall, the proposed framework offers a computationally efficient alternative to Monte Carlo simulations for designing energy- and cost-efficient quantized massive MIMO systems. Extending the analysis to coded transmission, spatially correlated or Rician channels, and wideband OFDM settings represents a natural direction for future work.

\appendix
\section{Quantized Channel Estimate}
\label{sec:Appendix_A}
In order to prove (\ref{eq:H_qLMMSE}), we need to substitute the quantized received signal vector $\mathbf Y_{pq}$ from (\ref{eq:Ypq}) for $\mathbf Y_p$ in (\ref{eq:H_LMMSE_eq}) and solve it again as follows: 
\begin{align}
\hat{\mathbf H}_{q} &=\underset{\mathbf W}{\arg\min} \ \mathbb{E} \Big\lbrace \parallel \mathbf H-(\alpha\sqrt{p_u}\  \mathbf H \mathbf S_p+\mathbf N_e)\mathbf W \parallel^2_F \Big\rbrace\nonumber\\
& =\underset{\mathbf W}{\arg\min} \ \mathbb{E} \Big\lbrace \Tr \Big[(\mathbf H-\alpha\sqrt{p_u}\  \mathbf H \mathbf S_p\mathbf W-\mathbf N_e\mathbf W )^H \nonumber\\
& \hspace{5em}\times (\mathbf H-\alpha\sqrt{p_u}\  \mathbf H \mathbf S_p\mathbf W-\mathbf N_e\mathbf W )\Big] \Big\rbrace  \nonumber\\
& =\underset{\mathbf W}{\arg\min} \Tr \Big[\mathbb{E} \Big\lbrace (\mathbf H-\alpha\sqrt{p_u}\  \mathbf H \mathbf S_p\mathbf W-\mathbf N_e\mathbf W )^H \nonumber\\
& \hspace{5em}\times (\mathbf H-\alpha\sqrt{p_u}\  \mathbf H \mathbf S_p\mathbf W-\mathbf N_e\mathbf W ) \Big\rbrace \Big] 
\end{align}
with the assumption of uncorrelated $\mathbf H$ and $\mathbf N_e$, and having $\mathbb{E}\lbrace\mathbf H^H \mathbf H\rbrace=N \mathbf I_K$ and $\mathbb{E}\lbrace\mathbf N_e^H \mathbf N_e\rbrace=N \sigma^2_{N_e}\mathbf I_K$, we have
\begin{align}
\hat{\mathbf H}_{q} &= \underset{\mathbf W}{\arg\min} \Tr \Big[ 
N\mathbf I_K - \alpha\sqrt{p_u}\, N \mathbf S_p \mathbf W 
- \alpha\sqrt{p_u}\, N \mathbf W^H \mathbf S_p^H \nonumber\\
&\quad + \alpha^2 p_u\, N \mathbf W^H \mathbf S_p^H \mathbf S_p \mathbf W 
+ N\sigma_{Ne}^2 \mathbf W^H \mathbf W \Big]
\end{align}
Assuming
\begin{align}
\mathbf C&= \Tr \Big[N\mathbf I_K-\alpha\sqrt{p_u}\  N \mathbf S_p\mathbf W-\alpha\sqrt{p_u}\  N \mathbf W^H\mathbf S_p^H\nonumber\\
&+\alpha^2 p_u\  N \mathbf W^H\mathbf S_p^H\mathbf S_p\mathbf W+N\sigma_{Ne}^2 \mathbf W^H\mathbf W\Big]\nonumber,
\end{align}
the optimal $\mathbf W$ can be found from the points where the first derivatives of $\mathbf C$ (with respect to $\mathbf W$) equal to zero, as below
\begin{align}
\frac{\partial\mathbf C}{\partial\mathbf W} &= -\alpha\sqrt{p_u}\  N \mathbf S_p^T+\alpha^2 p_u\  N \big(\mathbf W^H\mathbf S_p^H\mathbf S_p \big)^T \nonumber\\ 
& +N\sigma_{Ne}^2 \big(\mathbf W^H \big)^T =0
\label{eq:CW}
\end{align}
Solving (\ref{eq:CW}) and substituting $\sigma_{Ne}^2$ (see Appendix C), we have 	
\begin{align}
\mathbf W & =\frac{1}{\alpha}\sqrt{p_u}\ \mathbf S^H_p \big(p_u\mathbf S_p\mathbf S^H_p+[1+(\frac{1-\alpha}{\alpha})( K p_u+1)]\mathbf I_K\big)^{-1} \nonumber \\
& =\frac{1}{\alpha}\sqrt{p_u}\ \mathbf S^H_p \big(p_u\mathbf S_p\mathbf S^H_p+L(\alpha,p_u)\mathbf I_K\big)^{-1} 
\end{align}

\section{Channel Estimation Error}
\label{sec:Appendix_B}

To find the results in (\ref{eq:sigma_hq_hat_eq}), we need to use $\mathbf S_p\mathbf S_p^H=\tau\mathbf I_K$ in (\ref{eq:H_qLMMSE}) as follows:
\begin{align}
\hat{\mathbf H}_{q}&= \frac{1}{\alpha}\sqrt{p_u}\ \mathbf Y_{pq} \mathbf S^H_p \big(\tau p_u\mathbf I_K+L(\alpha,p_u)\mathbf I_K\big)^{-1}\nonumber \\ 
& =\frac{\sqrt{p_u}}{\alpha \big[\tau p_u +L(\alpha,p_u) \big]}\ \mathbf Y_{pq} \mathbf S^H_p 
\end{align}
substituting $\mathbf Y_{pq}$ from (\ref{eq:Ypq}) we have 
\begin{align}
\hat{\mathbf H}_{q} &= 
\frac{\sqrt{p_u}}{\alpha[\tau p_u + L(\alpha,p_u)]}
(\alpha\sqrt{p_u}\,\mathbf H \mathbf S_p + \mathbf N_e)\mathbf S_p^H \nonumber\\
&= \frac{\tau p_u}{[\tau p_u + L(\alpha,p_u)]}\,\mathbf H 
+ \frac{\sqrt{p_u}}{\alpha[\tau p_u + L(\alpha,p_u)]}\,\mathbf N_e \mathbf S_p^H
\end{align}
where the $(i,j)$ entry of $\hat{\mathbf H}_{q}$ is given by
\begin{align}
\hat{h}_{qij} &= \frac{\tau p_u}{\big[\tau p_u +L(\alpha,p_u) \big]}\ h_{ij} \nonumber \\
&+\frac{\sqrt{p_u}}{\alpha \big[\tau p_u +L(\alpha,p_u) \big]}\ \sum_{k=1}^{\tau} \mathbf N_e(i,k) \mathbf S^\ast_p(k,j)
\end{align}
Variance of $\hat{h}_{qij}$ can be calculated as below:
\begin{align}
&\mathbb{E} \Big\lbrace\vert \hat{h}_{qij} \vert^2\Big\rbrace =\mathbb{E} \Big\lbrace \Big\vert \frac{\tau p_u}{\big[\tau p_u +L(\alpha,p_u) \big]}\ h_{ij} \Big\vert^2\Big\rbrace \nonumber \\
&+\mathbb{E} \Big\lbrace \Big\vert \frac{\sqrt{p_u}}{\alpha \big[\tau p_u +L(\alpha,p_u) \big]}\ \sum_{k=1}^{\tau} \mathbf N_e(i,k) \mathbf S^\ast_p(k,j) \Big\vert^2\Big\rbrace \nonumber\\
&=\frac{(\tau p_u)^2}{\big[\tau p_u +L(\alpha,p_u) \big]^2}\mathbb{E} \Big\lbrace \vert \ h_{ij} \vert^2\Big\rbrace+\frac{p_u}{\alpha^2 \big[\tau p_u +L(\alpha,p_u) \big]^2} \nonumber \\
&\times\sum_{k=1}^{\tau}\mathbb{E} \Big\lbrace \Big\vert \ \mathbf N_e(i,k)  \Big\vert^2\Big\rbrace \mathbb{E} \Big\lbrace \Big\vert \ \mathbf S^\ast_p(k,j) \Big\vert^2\Big\rbrace \nonumber\\
&=\frac{(\tau p_u)^2}{\big[\tau p_u +L(\alpha,p_u) \big]^2} + \frac{p_u}{\alpha^2 \big[\tau p_u +L(\alpha,p_u) \big]^2}\sum_{k=1}^{\tau}\sigma_{Ne}^2 \cdot 1 \nonumber\\
&=\frac{(\tau p_u)^2}{\big[\tau p_u +L(\alpha,p_u) \big]^2}+\frac{\tau p_u \sigma_{Ne}^2}{\alpha^2 \big[\tau p_u +L(\alpha,p_u) \big]^2}
\end{align}
substituting $\sigma_{Ne}^2$ from equation (\ref{eq:sigma_Ne}), we can simply obtain the results in (\ref{eq:sigma_hq_hat_eq}).

\section{Variance of the noise matrix $\mathbf N_e$ }
\label{sec:Appendix_C}
The $(i,j)$ entry of the noise matrix $\mathbf N_e$ can be expressed as follows
\begin{equation}
\ n_e(i,j)=\alpha\ n_p (i,j)+\ n_{qp}(i,j)
\end{equation}
where the variance $\sigma_{Ne}^2$ is
\begin{equation}
\sigma_{Ne}^2= \mathbb{E}\big\lbrace\vert n_e(i,j)\vert^2\big\rbrace -\big\vert\mathbb{E}\big\lbrace n_e(i,j) \big\rbrace\big\vert^2
\end{equation}
The second term on the right hand side equals to zero. Because
\begin{align}
\mathbb{E}\big\lbrace n_e(i,j) \big\rbrace&= \alpha \mathbb{E}\big\lbrace n_p (i,j)\big\rbrace+\mathbb{E}\big\lbrace n_{qp}(i,j) \big\rbrace=0 
\end{align}
The other term can be simplified as follows:
\begin{align}
\mathbb{E}\big\lbrace\vert n_e(i,j)\vert^2\big\rbrace&=\mathbb{E}\big\lbrace n_e(i,j) \ n^\ast_e(i,j) \big\rbrace \nonumber\\ 
& =\alpha^2 \mathbb{E}\big\lbrace\vert n_p(i,j) \vert^2  \big\rbrace+\alpha\mathbb{E}\big\lbrace n_p(i,j)n^\ast_{qp}(i,j)\big\rbrace\nonumber\\
&+\alpha\mathbb{E}\big\lbrace n^\ast_p(i,j)n_{qp}(i,j)  \big\rbrace+\mathbb{E}\big\lbrace\vert n_{qp}(i,j) \vert^2  \big\rbrace\nonumber \\ 
&=\alpha^2+\alpha(1-\alpha)( K p_u+1)
\label{eq:sigma_Ne}
\end{align}
Note that samples of additive noise and quantization noise are uncorrelated in the above equation.

	\section{BER for $M$-QAM Modulations }
	\label{sec:Appendix_D}
To obtain (\ref{eq:berMQAM3}), we substitute (\ref{eq:PbAWGN}) into (\ref{eq:berMQAM1}), yielding:
\begin{align}
&\mathrm{BER}= \sum_{k=1}^{\log_2\sqrt{M}} \hspace{0.5em}\sum_{i=0}^{(1-2^{-k})\sqrt{M}-1}
\nonumber\\
&\Biggl\lbrace F(k,i) \int_0^\infty 
Q\!\Bigl(\sqrt{\tfrac{3(2i+1)^2}{M-1}\,\gamma_0 x}\Bigr)
f_{\chi_d^2}(x)\,dx\Biggr\rbrace
\label{eq:BER_MQAM_D}
\end{align}
To solve the above integral, we use the following  \cite{proakis2007digital} 
\begin{equation}
\int_{0}^\infty Q(\!\sqrt{cx}\!)f_{\chi_d^2}(x)\,dx = \!\Bigl(\tfrac{1-\mu}{2}\Bigr)^{\!\frac{d}{2}}\!\!\sum_{j=0}^{\frac{d}{2}-1}\!\binom{\frac{d}{2}-1+j}{j}\!\Bigl(\tfrac{1+\mu}{2}\Bigr)^j
\end{equation}
where
\begin{equation}
    \mu  = \sqrt{\frac{c}{2+c}}, \ \ \ \ c = \frac{3(2i+1)^2}{M-1}\gamma_0
\end{equation}
By substituting \(\gamma_0\) from (\ref{eq:gamma_q0}) into \(c\) and after simplifying, we obtain (\ref{eq:berMQAM3}), (\ref{eq:Bmu}) and (\ref{eq:mui}).

\bibliographystyle{elsarticle-num}  
\bibliography{ref}        

\end{document}